\documentstyle[prl,aps,twocolumn]{revtex}
\begin{document}
\draft
\title{Backscattering of  electromagnetic and gravitational
waves off Schwarzschild geometry  }
\author{Janusz Karkowski, Edward Malec}
\address{Institute of Physics, Jagiellonian University,
 30-059 Krak\'ow, Reymonta 4, Poland}
\author{and Zdobys\l aw \'Swierczy\'nski}
\address{Institute  of Physics,  Pedagogical University,
 30-  Krak\'ow,  Podchor\c a\.zych 1, Poland}

\maketitle

\begin{abstract}
This paper shows  that the backscattering of electromagnetic and
gravitational waves  can be dominant when the   radiation
is produced very close to a spherical black hole. Numerical investigation
shows that  almost  50 percent of the outgoing
quadrupole gravitational wave is backscattered, for a class of  initial
data. A similar analysis reveals at least
20 percent effect for a dipole electromagnetic radiation.
Numerical results confirm theoretical predictions that the backscatter of
short wavelength radiation is negligible.
In the long-radiation band is observed a rather weak  dependence on the
wavelength. Our studies base on the linear approximation. They can be of
relevance for the determination of the total energy of backscattering
tails and  quasinormal modes.

\end{abstract}

\date{ }

\section{ Introduction}

This paper is dedicated mainly to a  numerical
 investigation of the   backscattering 
(\cite{Hadamard} - \cite{Roszkowski:2001mp}) of waves in a
 Schwarzschild  spacetime. It supplements the earlier  analytic
studies   of  scalar \cite{Malec:1998au},
  electromagnetic (\cite{Malec:2000ji},
  \cite{Malec:2001ih}) and gravitational   fields \cite{Malec:2001iz}).    
 We  study
the propagation of the  electromagnetic and ( even-parity)
gravitational waves  in a  background Schwarzschild
spacetime. It is assumed  that initial data  describe
an isolated pulse (burst) of a (gravitational or electromagnetic)  wave.
  As in the    earlier studies (\cite{Malec:1998au}, \cite{Malec:2000ji} and
\cite{Malec:2001iz})
the strength of the backscattering is assessed by finding the fraction of
the initial burst energy that  will not reach a distant observer
in the main pulse. The numerical investigation,
 that is reported below, gives a quantitative evaluation of the effect.

Sec. II brings    an analytic estimate of the backscattering of the
electromagnetic fields.  The obtained  result is a substantial
improvement
of a former bound  \cite{Malec:2001ih}. Sec. III   is dedicated to the
presentation of
numerical results on the backscattering of electromagnetic fields. The
effect depends both on the relative
 width of the initial data and on the distance.
   Sec. IV  describes the propagation of quadrupole gravitational waves
in a Schwarzschild geometry. Relevant energy formulae  are
defined and analytic estimates are reminded.
Section V goes on with  numerical studies of the backscatter
of  gravitational  fields. Similarly as before, the
   backscattering happens to depend  on the  relative width of the initial
data
 and on the  distance.
In Sec. VI we briefly review the obtained results.

The space-time geometry  is defined  by
a Schwarzschildean   line element,
\begin{equation} ds^2 = - (1-{2m\over R})dt^2 +
{1\over 1-{2m\over R}} dR^2 +
R^2 d\Omega^2~,
\label{1}
\end{equation}
where $t$ is a time coordinate, $R$ is a radial
coordinate that coincides with the areal radius
and $d\Omega^2 = d\theta^2 + \sin^2\theta d\phi^2$
is the line element on the unit sphere, $0\le \phi < 2\pi $
and $0\le \theta \le \pi $.
Throughout this paper $G$, the Newtonian gravitational constant,
and $c$, the velocity of light are put equal to 1.
We define the Regge-Wheeler coordinate
\begin{equation}
r^*=R+2m\ln ({R\over 2m}-1)
\end{equation}
and $\eta_r\equiv 1- 2m/r$.

\section{Backscatter of electromagnetic waves: new analytic estimates}

We seek, following    \cite{Malec:2000ji},
a  solution $\Psi (r^*,t)$ in the form
\begin{equation}
\Psi =\tilde \Psi +\delta ,
\label{2.1}
\end{equation}
where $\delta $   satisfies the dipole equation
\begin{equation}
(-\partial_0^2 + \partial_{r^*}^2)\delta = \eta_R
\Biggl[ { 2\over R^2}
\delta + {6mf\over R^4}  \Biggr] .
\label{2.2}
\end{equation}
Here
\begin{equation}
\tilde \Psi (r^*,t)= \partial_t f(r^*-t) +{f(r^*-t)\over R}
\label{2.3}
\end{equation}
and $f$ is an arbitrary function with support in $(a, \infty )$.
 $f$  can be uniquely determined from initial data corresponding
 to an initially outgoing radiation.
 $\tilde \Psi $  solves Maxwell equations  in Minkowski
spacetime and it corresponds to the dipole radiation.
Initially $\delta =\partial_0\delta =0$.

The  energy  $E_R(t)$ of the electromagnetic field  $\Psi $
contained in the exterior of a sphere of the radius $R$ reads
\begin{equation}
E_R(t) =2\pi  \int_{R}^{\infty }dr
\Biggl(  {(\partial_0\Psi )^2\over \eta_r} + \eta_r
(\partial_r\Psi )^2+{2(\Psi )^2\over r^2}\Biggr) .
\label{2.4}
\end{equation}
 $E_a\equiv E_a(0)$ is the energy of the initial pulse.
Let  an outgoing  null geodesic $\tilde \Gamma_r$
originate from a point $(r,0)$
of the initial hypersurface.
In the Minkowski spacetime the outgoing radiation contained
outside $\tilde \Gamma_a$ does not leak inward and its
energy remains constant.
In a curved spacetime some  energy will be lost from the main stream
due to the  diffusion of the  radiation   through   $\tilde \Gamma_a$.

The  energy loss  is equal to a line integral  along      $\tilde
\Gamma_a$
(where $f=\tilde \Psi =0$),
\begin{eqnarray}
  \delta E_a\equiv  E_{a}- E_{\infty }=
2\pi \int_{a}^{\infty  } dr\Biggl[ \eta_r
\Bigl( {1\over \eta_r}(\partial_0+\partial_{r^*})\delta \Bigr)^2
  +{ 2 \delta^2\over r^2} \Biggl] .
\label{2.5}
\end{eqnarray}
An energy  $H(R,t)$ of  field $\delta$,
\begin{equation}
H(R,t) = \int_{R}^{\infty }dr
\Bigl(  {(\partial_0\delta )^2\over \eta_r} + \eta_r
(\partial_r\delta)^2+(\delta)^2{ 2\over r^2}\Bigr) ,
\label{2.6}
\end{equation}
satisfies the "conservation" equation \cite{Malec:2000ji}
\begin{eqnarray}
&&
(\partial_t+\partial_r^*)H(R,t)= \nonumber\\ &&
-\eta_R\Biggl[ \eta_R
\Bigl( {\partial_0\delta \over \eta_R} +
\partial_R\delta
 \Bigr)^2 +{ 2\over R^2}\delta^2 \Biggr] -
12m \int_R^{\infty }dr\partial_0\delta
{ f\over r^4}
\label{2.7}
\end{eqnarray}
The integration of  (\ref{2.7}) along  $\tilde \Gamma_a$ yields
\begin{equation}
H(\infty ,\infty )-H(a,0)=-{\delta E_a \over 2\pi }-
12m\int_0^{\infty }dt \int_{a_t}^{\infty }dr\partial_0\delta
{ f\over r^4} ;
\label{2.8}
\end{equation}
here $(a_t,t)\in \tilde \Gamma_a$ and $H(\infty ,\infty )$  is the
asymptotic
 energy of the field  $\delta $.
Eq. (\ref{2.8}) implies (taking into account that
 $H(a,0)=0$, due to the vanishing
of initial data of the field $\delta $) that
\begin{eqnarray}
\delta E_a &\le &  -2\pi H(\infty ,\infty )  +
24\pi m|\int_0^{\infty }dt \int_{a_t}^{\infty }dr\partial_0\delta { f\over
r^4}|
\le   \nonumber \\ &&
24\pi m|\int_0^{\infty }dt \int_{a_t}^{\infty }dr\partial_0\delta { f\over
r^4}|.
\label{2.9}
\end{eqnarray}
This integral is bounded from above by
\begin{equation}
 24\pi m
\int_0^{\infty }dt \sqrt{H(t)} \Biggl( \int_{a_t}^{\infty }dr {
f^2\eta_r
\over r^8}\Biggr)^{1/2}.
\label{2.10}
\end{equation}
On the other hand, Eq. (\ref{2.7}) implies
\begin{eqnarray}
\sqrt{ H(t)}\le 6m \int_0^tdt
\Biggl( \int_{a_t}^{\infty }dr { f^2\eta_r
\over r^8}\Biggr)^{1/2};
\label{2.11}
\end{eqnarray}
thus
\begin{eqnarray}
&& 24\pi m
\int_0^{\infty }dt \sqrt{H(t)} \Biggl( \int_{a_t}^{\infty }dr {
f^2\eta_r
\over r^8}\Biggr)^{1/2}\le  \nonumber \\
&& 144 \pi m^2  \int_0^{\infty }dt \int_0^tds
\Biggl( \int_{a_s}^{\infty }dr { f^2\eta_r
\over r^8}\Biggr)^{1/2} \nonumber \\
&&\Biggl( \int_{a_t}^{\infty }dr { f^2\eta_r
\over r^8}\Biggr)^{1/2}.
\label{2.12a}
\end{eqnarray}
The integrand of the external integral with respect "t"
is equal to the directional derivative
${d\over dt}\Biggl[ \int_0^tds
\Bigl( \int_{a_s}^{\infty }dr { f^2\eta_r
\over r^8}\Bigr)^{1/2}\Biggr]^2$
and the last line of the preceding equation is equal to
\begin{eqnarray}
&& 72 \pi m^2  \int_0^{\infty }dt {d\over dt}\Biggl[  \int_0^tds
\Biggl( \int_{a_s}^{\infty }dr { f^2\eta_r
\over r^8}\Biggr)^{1/2}\Biggr]^2=  \nonumber \\
&&72 \pi m^2   \Biggl[  \int_0^{\infty }dt
\Biggl( \int_{a_t}^{\infty }dr { f^2\eta_r
\over r^8}\Biggr)^{1/2}\Biggr]^2
\label{2.12}
\end{eqnarray}
Therefore, (\ref{2.9}-\ref{2.12}) imply
\begin{equation}
\delta E_a\le   36m^2\Biggl[ \int_0^{\infty }dt\Bigl(
\int_{a_t}^{\infty }dr
{ 2\pi f^2\eta_r\over r^8} \Bigr)^{1/2}\Biggr]^2 .
\label{2.13}
\end{equation}
One can show, using reasoning as in  \cite{Malec:2001ih},
that on the initial hypersurface $t=0$
\begin{equation}
|{f(R, t=0)\over R-2m}|=|\int_a^R\partial_r{f\over r-2m}|=
|-\int_a^Rdr {r\tilde \Psi \over (r-2m)^2}|;
\label{2.13a}
\end{equation}
this can be  written, using the Schwarz inequality, as
\begin{equation}
|{f(R, t=0)\over R-2m}|\le
\Biggl(  \int_a^Rdr \bigl( {\tilde \Psi \over r}\bigr)^2
 \int_a^Rdr{r^4\over (r-2m)^4}\Biggr)^{1/2}.
\label{2.13b}
\end{equation}
Notice that   $\int_a^Rdr \bigl( {\tilde \Psi \over r}\bigr)^2\le
E_a/(4\pi )$.
 From (\ref{2.13b})  follows
\begin{eqnarray}
{4\pi  f^2\over r^2}\le E_a a \eta_r^2F(\tilde m,y),
\label{2.14.0}
\end{eqnarray}
where
\begin{eqnarray}
F(\tilde m,y) &\equiv &
 y-1+{16\tilde m^4\over 3 (-y+2\tilde m)^3}-
{16\tilde m^4\over 3 (-1+2\tilde m)^3}-
\nonumber\\
&&{16\tilde m^3\over
(-y+2\tilde m)^2}
+{16\tilde m^3\over (-1+2\tilde m)^2}+
\nonumber\\
&& {24\tilde m^2\over -y+2\tilde m}  -{24\tilde m^2\over -1+2\tilde m }
+\nonumber\\
&&8\tilde m \ln {y-2\tilde m\over 1-2\tilde m}.
\label{2.14}
\end{eqnarray}
Here  $\tilde m\equiv m/a$ and $y=R/a$.
Inequality (\ref{2.14}) constitutes a refined version of Eq. (13)
in \cite{Malec:2001ih}.

Let $(R,t)\in \tilde \Gamma_r$. The right hand side of (\ref{2.14})
is an increasing function of $r$; since $R>r$ and $f$ is constant
along $\tilde \Gamma_r$, one obtains
\begin{equation}
{4\pi f^2(R,t)\over R^2}\le E_a a \eta^2_RF(\tilde m,R/a),
\label{2.14a}
\end{equation}
Define $x\equiv a_t/a$.
The integrand of the
 right hand side of (\ref{2.13}) is bounded, taking into account
(\ref{2.14a}):
\begin{equation}
 2\pi \int_{a_t}^{\infty } \eta_r{f^2\over r^8}\le
 {E_a a\over 2}\int_{a_t}^{\infty }
 {\eta^3_rF(\tilde m,r/a)\over r^6} dr ={E_a\over 2a^4}G(\tilde m,x)  .
\label{2.15}
\end{equation}
Here $G(\tilde m,x)=\int_x^{\infty }dx(x-2\tilde m)^3F(\tilde m, x)/x^9$;  
  the rather lengthy integration can be done explicitly with the use of Mapple. 
The result is too long to be presented here..

The insertion of (\ref{2.15}) into   (\ref{2.13})
leaves us with
\begin{equation}
\delta E_a\le  {18m^2E_a\over a^2}\Biggl[ \int_1^{\infty }{dx\over
1-{2\tilde m \over x}}
\sqrt{G(\tilde m,x)}\Biggr]^2.
\label{2.16}
\end{equation}
Here we replaced the time variable $t$ by the radial variable $r$;
notice that along $\tilde \Gamma_a$ one has $dt =dr/\eta_r$.
The use of the Schwarz inequality allows one to bound (\ref{2.16})
from above by
\begin{equation}
\delta E_a\le {18m^2E_a\over a^2} \int_1^{\infty }dx G(\tilde m,x) x^2
 \int_1^{\infty }dx {1\over  (x-2m)^2}  .
\label{2.17}
\end{equation}
That implies the following estimate for the ratio of the diffused energy
by the initial energy of the pulse of radiation
\begin{eqnarray}
{\delta E_a \over E_a}\le 4.5\Bigl( {2m\over a}\Bigr)^2 C(\tilde m)
\label{2.18}
\end{eqnarray}
where
\begin{eqnarray}
C(\tilde m)&=& {-1\over 10080((-1+2\tilde m)^4\tilde m^4)}\Biggl(
-2760\tilde m^5+828\tilde m^4+\nonumber \\
&& 44\tilde m^3+2352\tilde m^6+2016\tilde m^4\ln (1-2\tilde m)+
\nonumber \\
&&2688\tilde
m^6\ln (1-2\tilde m)- 4032\tilde m^5\ln (1-2\tilde m)-\nonumber \\
&&360\tilde m^3\ln (1-2\tilde m)+6\tilde
m+36\tilde m^2\ln (1-2\tilde m)\nonumber\\
&&-30\tilde m^2+3\ln (1-2\tilde m)-\nonumber \\
&&18\tilde m\ln (1-2\tilde m)\Biggr)
\label{2.19}
\end{eqnarray}
This result can be improved in the case
when the initial pulse is
located outside the  sphere $a=10m/3$. One finds, eliminating the
time derivatives of $\tilde \Psi $, that the energy can be written as
\begin{eqnarray}
&&E_a(t) =4\pi  \int_{a}^{\infty }dr
\Biggl(   \eta_r(\partial_r\tilde \Psi )^2+{2(\tilde \Psi )^2\over r^2}+
 \nonumber \\
&&{20mf^2\over r^5}(0.9-{3m\over r})\Biggr) .
\label{2.21}
\end{eqnarray}
Thus $\int_a^{\infty }\tilde \Psi^2/r^2dr \le  E_a/(8\pi )$ if $a>10m/3$.
As a consequence the factor in the inequality  (\ref{2.14a}) halves
and, finally,  the formula (\ref{2.18}) can be replaced  by
\begin{eqnarray}
{\delta E_a \over E_a}\le  2.25\Bigl( {2m\over a}\Bigr)^2 C(\tilde m)
\label{2.21a}
\end{eqnarray}
A more careful estimate, in which the $\partial_r\tilde \Psi $-related term
is taken into account, gives even a stronger result, with
\begin{eqnarray}
{\delta E_a \over E_a}\le  {2.25\over 1+0.125\eta_a}
\Bigl( {2m\over a}\Bigr)^2 C(\tilde m) .
\label{2.21b}
\end{eqnarray}
In the limit  $\tilde m\rightarrow 0$ one arrives at
\begin{eqnarray}
{\delta E_a \over E_a}&\le & 0.3\Bigl( {2m\over a}\Bigr)^2;
\label{2.20}
\end{eqnarray}
that significantly improves the former result proven in \cite{Malec:2001ih}.
The coefficient $C(m)$ diverges at $\tilde m=0.5$ but
it depends rather weakly on
 $\tilde m=m/a$ in the range $(0,0.25)$.
For instance, for  $m/a=0.1$ and $m/a=0.25$, one has
$2.25C(\tilde m)/(1+0.125\eta_a)\approx 0.39$ and
$2.25C(\tilde m)/1+0.125\eta_a)\approx 1.4$, respectively.

It can be  of interest to consider the case of initial
data of compact support $(a,b)$. The simplest estimate can be obtained
as follows.  It is easy to see that  one has
\begin{equation}
{4\pi f^2(R,t)\over R^2}\le E_a a \eta^2_RF(\tilde m,b/a),
\label{2.14ab}
\end{equation}
instead of (\ref{2.14a}). The insertion of (\ref{2.14ab}) into Eq. (\ref{2.13})
yields, after a calculation analogous to that performed above,
\begin{eqnarray}
{\delta E_a \over E_a}\le  0.45\Bigl( {2m\over a}\Bigr)^2 
F(\tilde m, b/a) \Bigl( 1-{  m\over a(a-2  m)}({3\over 7}-
{\tilde m \over 4})\Bigr) .
\label{2.21ab}
\end{eqnarray}
This is valid also inside the photon sphere. In the case  
when $(b-a)/a<<\eta_a^3$, one can clearly see that the
 backscatter is negligible.

A simpler and much stronger estimate can be obtained, assuming that
$2m<<a$ and the support of initial data satisfies the condition
$b-a<<a$.  In this case one has from (\ref{2.13b})

\begin{equation}
|{f(R, t=0)\over R}|\le
\Biggl(  (b-a) \int_a^Rdr \bigl( {\tilde \Psi \over r}\bigr)^2
  \Biggr)^{1/2}.
 \label{22b}
\end{equation}
Here $\Psi (a) =0$ and one can show, employing the same approach as
before,
that
\begin{eqnarray}
|\Psi (R)|&=&
|\int_a^Rdr \partial_r\Psi |=|\int_b^Rdr \partial_r\Psi |\le \nonumber\\
&& \sqrt{E_a\over 2\pi }\min \Bigl( \sqrt{R-a},  \sqrt{b-R}\Bigr) .
\label{22c}
\end{eqnarray}
Thus
 \begin{equation}
{4\pi f^2 (R)\over R}\le {E_a\over 4} b^2(1-{a\over b})^3.
\label{2.23}
\end{equation}
The insertion of (\ref{2.23}) into (\ref{2.13}) yields (notice that the
spatial
integration now extends from $a_t$ to $b_t$ and that $(a_t-b_t)/a\approx
(b-a)/a$)
\begin{eqnarray}
{\delta E_a \over E_a}\le 0.5 \Bigl( {m\over a}\Bigr)^2  \Bigl( {b-a\over
a}
\Bigr)^4.
\label{2.24}
\end{eqnarray}

\section{Numerical results: electromagnetic fields}

This section reports numerical results on the backscatter. We begin with
 demonstrating that the effect can be strong and then discuss
qualitatively
 its wavelength dependence.

We choose initial data   generated by the function
\begin{eqnarray}
\partial_tf&=&(R-a)^{2.01} \exp \bigl( -w(R-a)^2\bigr) ,~~~~~R\ge
a;
\nonumber \\
&&=0~~~~~~~~~~~~~~~~~~2m\le R<a.
\label{3.1}
\end{eqnarray}
There are two free parameters,  $a$ and $w$.  The exponent  $2.01$
 guarantees that the energy density (that depends, in particular,
on    $\partial_R^2f$) vanishes at $a$.

The initial data have a noncompact domain, but  they
are very small outside   a compact support.
In particular, the
projected one-dimensional initial energy density
(the integrand of  Eq. (\ref{2.4})) is practically of compact support.
 On the other hand, the function $f$ is obtained from Eq. (\ref{3.1})
 by integration
and it is  constant asymptotically.
Since $f$ enters the evolution equation (\ref{2.2}),
it is responsible for the backscattering; its asymptotic constancy ensures
that the effect is relatively strong. In what follows the mass $m$ is
assumed
to be 1. The backscattered energy is obtained by a numerical approximation
of the line integral (\ref{2.5}).

 We integrated Eq. (\ref{2.2}) for $a=2, 4$ and  $20$.
Fixing the parameter $a$, calculations were
performed for a  number of
$w's$. The results seem to imply  that $\delta E_a/E_a $ has
a single maximum at some $w_*$, with $\delta E_a/E_a $ being almost
constant in some vicinity of $w_*$.
The limitation inherent to  the
numerics does allow us  to find  only an approximate value of $w_*$;
that value will be  called as a "maximal" point and the corresponding
value of  $\delta E_a/E_a $ will be referred to as a "maximum".
That terminology will be kept in the rest of Sec. III as well as  in
Sec. V.

\subsection{The maximal backscattering  }

The strongest backscattering effect is found with
  the choice of parameters $a=2$ and $w=5\times 10^6$.
  The backscattered energy has been obtained by the integration
of  (\ref{2.5}) along $\tilde \Gamma_{2.0001}$ \cite{explanation}.
 Under these conditions the  value of the ratio $\delta E_a/E_a $
approximates $ 20.5 \% $.
The backscatter is relatively insensitive on the choice of $w$.
 Fig. 1a)   shows the initial energy density for
 various  values of  $w$ as well as the strength of the backscatter as
measured by the ratio  $\delta E_a/E_a$.

\subsection{Resonant type initial data and relative-width dependence}

 Figs. 1b) and 1c) show the behaviour of the initial energy density
that corresponds to  $a=4$ and $a=20$, respectively,
and to   a number of values of $w$.
  A number assigned to  a curve gives the ratio of the backscattered
energy.
It is obvious  that in these numerical examples the  strength of the
backscatter is
correlated with the width of the support $\Delta $ of initial data.
The relative width $\Delta /a$ of the energy density is related to its
spectral
composition; that suggests that the effect depends on the wavelength  of
the initial radiation.

Numerical investigation confirms a theoretically derived conclusion that
in the limit
$\Delta /a\rightarrow 0$ there is no backscatter (\cite{Malec:2000ji};
see also Eq.  (\ref{2.24})). In our examples small values of the
ratio $\Delta /a$ correspond to $w>>1$.  We found that the ratio  $\delta
E_a/E_a$
monotonically decreases with $w$, if $w$ is large enough. In order to
demonstrate
how dramatic the changes can be, we compare  data concerning the case
$a=4$.
 If $w=10^3$, then $\delta E_a/E_a =4.5\times 10^{-8}$;  that is rougly
$10^4$ less
than  for $w=1$, when  $\delta E_a/E_a =1.9\times 10^{-4}$.

The values $w=0.1$ (in the case of  $a=4m$) and  $w=0.001$ (in the case of
$a=20m$)
correspond to "resonant" initial data. That "resonant" behaviour is rather
weak; for
 instance in the case of $a=4$ the change of  $w$ by a factor
of 10  from the maximal case
  results in a relatively small,  less than 50\%, change of the efficiency
factor
$\delta E_a/E_a$ of the  backscattering. In general, we can
say that the "resonant" width of the initial energy density (which gives
also
the order of the "resonant" wavelength of the radiation, if $a>>2$)
 is of the order of the distance
of the radiating source from the black hole.

 Fig. 1a) clearly   shows, in contrast with the preceding cases,
 that the strength of backscatter is rather weakly
correlated with the width of the support of initial data.
The reason for that is  that (for this set of configurations),
the smaller is the width, the smaller is the
distance from the horizon (and thus  the redshift increases).
There are two  competing effects  that work each against  the other.

In those cases that have been studied in our paper, the  backscattering
 is strongest at  $a=2$ and $w=5000000$, when $\delta
 E_{2.0001}/E_{2.0001}=20.5\% $.
The resonant efficiency drops then  by a factor of 500 when passing to
another  resonant case at $a=4$ and $w=0.1$:
$\delta E_4/E_4=.045\% $. This dramatic change is expected, since the
backscattering is strongest within the sphere $R=3m$. The
largest value of $\delta E_{20}/E_{20}$ (and $w=0.001$) is obtained
 for  $w=0.001$ and it  reads $0.00085\% $.
 Thus the increase of $a$ from 4 to 20 again results
in a roughly 50-fold decrease of the resonant efficiencies, which is
somewhat
quicker than fall-off suggested by the  analytic estimate (\ref{2.18}).
 
\subsection{On the comparison of analytic and numerical results}

Let us comment on the comparison of the numerical data with the
  analytic estimates of the preceding section. The  inequality
(\ref{2.18},  valid generally, is not sharp.
In the case of resonant initial data, for instance, it gives a nontrivial
information only  for $a=4m$ and  $a= 20m$  (Figs. 1b and
1c); the numerically obtained
value of $\delta E_a/E_a$ is   by three orders   smaller than the
analytic bounds. The   formula (\ref{2.24}), that is valid at large
distances
and for initial data with small relative widths,  should be
more efficient. The use of (\ref{2.24}) is not allowed, strictly saying,
 in our case  since  the numerical examples  do not satisfy the
required  assumptions concerning the compactness of initial data.
In the limit of $\Delta /a\rightarrow 0$, however, the initial
data have "almost" compact support and   in this context
  it is interesting that (\ref{2.24}) gives predictions
 that are comparable with numerical ones, in the case of initial data
 with narrow support.  We will study elsewhere the question whether (and
under
which conditions) the analytic criteria are strict.

\section{Propagation of  gravitational waves}

  We will search the solution of the Zerilli equation \cite{Zerilli}
  in the form
\begin{equation}
\Psi =\tilde \Psi +\delta ,
\label{5}
\end{equation}
where $\delta $ is an  unknown function satisfying the
(quadrupole)  equation
\begin{eqnarray}
&&(-\partial_t^2 + \partial_{r^*}^2)\delta  = V\delta +
(V-6{\eta_R^2\over R^2})\Biggl( \Psi_0+{\Psi_1\over R}+{\Psi_2\over
R^2}\Biggr) +
\nonumber\\
&&  { 2m\eta_R\over R^4}\Biggl[ -3\Psi_1 +2{\Psi_2\over R}  \Biggr] .
\label{11a}
\end{eqnarray}
Here   
\begin{equation}
V(R)=6\eta_R^2  {1\over R^2} +
\eta_R{63m^2(1+{m\over R})\over 2R^4(1+{3m\over 2R})^2}
\label{3a}
\end{equation}
and $ \Psi_i(r^*-t)$, $i=0,1,2$, are functions
that satisfy the  relations
\begin{eqnarray}
&&\partial_t  \Psi_1= 3  \Psi_0\nonumber\\ &&
\partial_t   \Psi_2= \Psi_1 - m\partial_t \Psi_1.
\label{4a}
\end{eqnarray}
The combination
\begin{equation}
\tilde \Psi \equiv  \Psi_0(r^*-t)+ {  \Psi_1(r^*-t)\over R} +
{  \Psi_2(r^*-t)\over R^2},
\label{4aa}
\end{equation}
which  represents a purely outgoing radiation,
solves the Zerilli equation in  Minkowski space-time (m=0).
We choose $\Psi = \tilde \Psi $, $\partial_t \Psi =   \partial_t\tilde
\Psi $,
which implies $\delta =\partial_t\delta =0$ at $t=0$.

The initial  energy density multiplied by $R^2$ reads
\begin{equation}
 \rho =\bigl( (\partial_t\Psi )^2 + (\partial_{r^*}\Psi )^2
+V\Psi^2\bigr) /\eta_R  .
\label{rho}
\end{equation}
The initial data  are assumed to be smooth and to be nonzero
outside a sphere of a radius $a >2m$. Thus
$\rho $ is  smooth and it vanishes  on the boundary $a$.
The energy content inside  a part of a  Cauchy hypersurface $\Sigma_t$
that is exterior to a ball of a radius $R$ can be defined as
\begin{equation}
E(R,t)\equiv \int_{R}^{\infty }dr \rho (r,t).
\label{29a}
\end{equation}
 We omit a  normalization constant in the definition of the energy
$E(R,t)$,
since we  will be interested only in the relative efficiency
of the backscatter and the normalization factor cancels out.
The total initial energy corresponding to the  hitherto defined
initial data will be written as $E_a$.

The  energy loss, that is the amount of energy that diffused
inward $\tilde \Gamma_a$ is equal to a line integral  along
$\tilde \Gamma_a$,
\begin{eqnarray}
&& \delta E_a \equiv  E(a,0)- E_{\infty }=
 \nonumber\\ &&
 \int_{a}^{\infty  } dr\Biggl[ \eta_r
  \Bigl( {1\over \eta_R }(\partial_t+\partial_{r^*})\delta (R,t)\Bigr)^2
  +{ V \delta^2\over \eta_r } \Biggl] .
\label{29}
\end{eqnarray}
It is necessary to point out that in the case of the initial point $R_0>a$
the result would be more complicated; the differentiation of the energy
along $\tilde \Gamma_{R_0}$
would depend also on $\Psi_0, \Psi_1$ and $\Psi_2$. If, however, the
outgoing null geodesics is $\tilde \Gamma_{a}$, then it starts from $a$
where   $\Psi_0, \Psi_1$ and $\Psi_2$  do vanish. Since these functions
depend on the difference $r^*-t$, their values along outgoing geodesics
are
constant, and that  allows one to conclude that they vanish at
 $\tilde \Gamma_a$.

   The  fraction
 of the    energy that could diffuse through
 the null cone $\tilde \Gamma_a$ satisfies
\cite{Malec:2001iz}

{\bf Theorem.}   The efficiency of the backscattering,
   $\delta E_a/E_a  $
satisfies the inequality
\begin{eqnarray}
&&  {\delta E_a\over E_a} \le  54.5 \times \Bigl( {2m\over a}\Bigr)^2
  +O(m^3/a^3).
\label{27}
\end{eqnarray}
In the case of compact initial  pulses one has  \cite{Malec:2001iz}, assuming
$(b-a)/a<<1$
and $m/a<<1$,
\begin{eqnarray}
  {\delta E_a\over E_a} \le
    \Bigl( { 2m \over a }\Bigr)^2\Bigl( {b-a\over a}\Bigr)^4.
\label{5.8}
\end{eqnarray}

\section{Numerical results:  gravitational  fields}

Below we shall describe numerical results on the backscattering
of even-parity gravitational waves. As in the case of electromagnetic
fields,  initial data are found that give rise to the strongest
effect and  then   the  wavelength dependence is discussed.
Finally,   the  efficiency coefficient $\delta  E_a/E_a$
corresponding to  resonant initial data is shown to decrease
with the increase of a distance.

We choose initial data   generated by the function
\begin{eqnarray}
 \Psi_1&=&(R-a)^{2.01} \exp \bigl( -w(R-a)^2\bigr), ~~~~R\ge a;
\nonumber \\
&&=0~~~~~~~~~~~~~~~~~~~~~~~2m\le R<a.
\label{5.1}
\end{eqnarray}
The exponent is taken to be $2.01$, in order to guarantee that the initial
energy density (that involves second derivatives of $\Psi_1$) vanishes
at $R=a$.
There are two parameters,  $a$ and $w$.
The initial data are  of noncompact
support  but from the numerical point of
view   they  are zero outside a compact set.
The one-dimensional projected initial energy density   (\ref{rho})
becomes negligibly small outside a compact support.
  The function $\Psi_2$ is obtained   by integration (compare
(\ref{5.1}) and (\ref{4a}))
and it is  constant asymptotically.
$\Psi_1 $ and $\Psi_2$ appear in the evolution equation (\ref{2.2}).
They probably generate the dominant contribution to the
 backscattering and the  constancy of $\Psi_2$ at spatial
 infinity can ensure the
best conditions for having the strongest effect.
 We put  $m=1$.
As in Sec. III,  Eq. (\ref{2.2}) is integrated
for $a=2, 4$ and  $20$. The initial energy is calculated from
the formula  (\ref{29a}) and the backscattered energy is found numerically
from
Eq. (\ref{29}). For each fixed parameter
$a$ is determined a value $w$ at which the factor
$\delta E_a/E_a$  is the largest one (but see the explanation in
Sec. III).

\subsection{The maximal  backscattering}

The highest ratio $\delta E_a/E_a$ is found
  for  parameters $a=2$ and $w= 10^4$. The
  backscattered energy was calculated along $\tilde \Gamma_{2.001}$
 \cite{explanation}
  and then   it was noticed that
  $\delta E_{2.001}/E_{2.001}  \approx 47 \% $. Let us point out that
  this  exceeds by a factor
of 10 a   prediction made in  \cite{Price:1993pi}.
The backscatter does not depend strongly on $w$.
 Fig. 2a)   shows the initial energy density for
 various   choices of $w$. A number assigned to a particular curve
shows  the strength of the backscatter - the
corresponding value of  $\delta E_{2.001}/E_{2.001}$.
 
\subsection{ Resonant backscatter, relative width dependence and
analytic estimates}

Figs. 2b) and 2c) show the    initial energy density
at  $a=4$ and $a=20$, for a selection of values of $w$,
with numbers showing values of the  fraction of the  backscattered energy
that are associated with particular initial data.
Figs. 2b) and 2c) reveal  that the  strength of the backscatter is
correlated with the relative width $\Delta /a$ of the support of initial
data.
 In contrast with that, Fig. 2a)  shows a rather weak
dependence  of   $\delta E_a/E_a$ with the
 (initial) relative width.
The explanation for that anomalous behaviour  is as in the case of
Fig. 1a - that
there do appear contradictory effects - the diminishing of the
relative width (with $w$ being increased) goes in pair  with
the decrease of the distance from the horizon.

The efficiency of the backscatter quickly diminishes with the decrease
of $\Delta /a$, as theoretically predicted \cite{Malec:2001iz}. In our
numerical
examples the ratio  $\Delta /a$ becomes smaller if $w$ increases. It was
found,
for sufficiently  large values of $w$,
that  $\delta E_a/E_a$ monotonically decreases with $w$. For instance, if
$a=4m$,
then the change of $w$ from $10^{-1}$ to $4\times 10^3$ results in the
decrease of $\delta E_a/E_a$ by a factor of $10^8$: from $3.7\times
10^{-3}$ to
$1.35\times 10^{-11}$.

The values  $w=0.01$ (in the case of  $a=4m$) and  $w=0.0001$
 (in the case of $a=20m$)
correspond to the  maxima  of the  factor  $\delta E_a/E_a$.
Again, as in sec. III,  "resonant" peaks  are rather mild.
  In the case of $a=4$ the change from   $w=0.1$ to $w=0.001$
changes the efficiency from  $\delta E_4/E_4=0.37 \% $ to
$\delta E_4/E_4=0.4 \% $, whereas the maximum is $0.65 \% $ at $w=0.01$.
 Notice, however, that the "resonant" width of the initial energy density
(which gives also
the order of the "resonant" length of the radiation, if $a>>2m$)
 is now  much bigger (by one order)
than   in the case of the electromagnetic dipole radiation.

The  backscattering
 is strongest at  $a=2$ and $w=10000$, when $\delta
 E_{2.001}/E_{2.001}=47\% $.
The efficiency  decreases quickly,  circa  70-fold,
when the pulse is moved outside the photonic sphere
 to  $a=4$. Then   the  resonant value
$\delta E_4/E_4=.65\% $ corresponds to  $w=0.01$.  At $a=20$
and $w=0.0001$ we found $\delta E_{20}/E_{20}=.01\% $.
Thus the increase of $a$ from 4 to 20   results
in  65-fold decrease of the efficiency at resonant cases, which is
more  than the  fall-off suggested by the analytic bound (\ref{27}).

The comparison of the numerical data with the
analytic estimates   (\ref{27}) and (\ref{5.8}) yields conclusions similar
to those made in the electromagnetic case.
Again, the width independent criterion (\ref{27}) is rather imprecise.
 In the case of  $a=20$ (Fig. 2c)  the numerical
value of $\delta E_a/E_a$ is circa 3 orders   smaller than that predicted
 analytically.  The  prediction of
the bound (\ref{5.8}), that is valid for $a>>m$ and
$b-a<<a$   can be   more precise.

\section{Conclusions}

There took place  a debate, about decade ago,  on the quantitative
evaluation of the backscattering of gravitational waves
(see \cite{Price:1993pi} and references therein).  While one of the proponents
 was originally in favour of a very strong damping of the radiation, that
was
implied by the backscattering, the final conclusion was that
the effect is weak  and only a small fraction (of the order of a few
percents)
 of the  long-wave band of the  gravitational
radiation   can be backscattered \cite{Price:1993pi}.
In the light  of that the main results of this paper
  come as a surprise. It is quite likely  that the
 ratio of almost 50$\% $ of the backscattered  gravitational
 quadrupole radiation, that we find here,
 can be  improved for more suitably chosen initial data.
 The same can be said about the backscatter of
electromagnetic waves, where we established that this effect exceeds
20 $\% $, for the dipole radiation.

The numerical examples of Sects. III and V confirm theoretical predictions
that
 if the relative width of the initial pulse  tends to zero
then the effect becomes negligible. This can be translated
(using the so-called similarity theorem of Fourier transforms)
 into dependence on the asymptotic
 wavelength (or frequency)  of the radiation \cite{Malec:2001ih}.
  The simplest argument  would  invoke to  the Heisenberg principle, which
   clearly implies  that
the compression of the support of a function leads to the
 increase of the frequency scale in its Fourier transform.
  The analytic proofs
can be deduced from   formula \ref{2.24})
and (\ref{5.8}) of this paper.
On the other hand, in the "resonance" regime the   dependence on the
relative width
(or, accepting the preceding arguments, on
  the  wavelength)  is rather weak, according to numerical data
of Secs. III and V. That seems to be in a sharp contrast with what is
known in the
case of stationary processes \cite{kto?}.

Let us  point out the difference between the manifestations
of the gravitational redshift and of the backscattering. The gravitational
redshift is
responsible for the weakening of the intensity of a radiation, but -
barring  backscatter -
 all of the  initial energy  eventually reaches an asymptotic observer. In
the
limit of the geometric optics the gravitational redshift is the only
phenomenon that can be
observed.  The backscattering in turn is responsible for the loss of
energy,
and it may be important in the case of a low-frequency radiation.

We observed in Secs III and V that the strongest  effect takes place very
close to the horizon.
That reflects well the known fact that the bulk of the backscatter
happens
inside the marginally stable photon sphere $R=3m$.  As a consequence,
 the transport of an initial
pulse from a location close to the horizon to a point behind $3m$ must
be associated with  a significant decrease in
the efficiency. That was in fact observed in our numerical examples,
both for gravitational and electromagnetic waves.
This  fall-off with a distance can be faster than
 $1/R^2$, which is typical for our  analytic estimates.

 The comparison of the analytic and numerical results
suggests that   analytic bounds ((\ref{2.18})and (\ref{27}))   are
  not strict; their predictions are bigger from our numerical data
 by  two - three orders. That gap can be made much smaller by a  more
 suitable choice of initial data, but we do not expect that it could be
 nullified.  
 The  estimates (\ref{2.24}) and (\ref{5.8})
 that are specialized to the case
of small relative widths  are expected to be much sharper.
 The analytic bounds on the efficiency of the backscatter clearly show
that the effect is negligible if the initial burst is  located far from
the horizon, irrespective of the detailed character of initial data.

We would like to point out a new important  application of the methods
developed in this paper. Namely, the backscattered (or diffused, in the
terminology of \cite{Malec:2000ji}) energy bounds from above the total energy
of  tails and quasinormal modes of black holes and of a class of quasinormal
modes ("w-"modes) of neutron stars (\cite{Nollert:1999ys}, \cite{Kokkotas:1999bd}).
Therefore the
present methods can be used as the first step in order
to estimate the fraction of the energy of  gravitational waves that can be
carried by   quasinormal modes in spherically symmetric spacetimes.
That would give an indication of what can be expected in more
realistic situations.

{\bf Acknowledgements.}  
One of us (EM) thanks Bernd Schmidt for  a discussion on quasinormal modes.
This work has been supported in part  by
the KBN grant 2 PO3B 010 16. Z. \'Swierczy\'nski thanks  the
Cracow Pedagogical University for the  research grant.


\begin{references}
\bibitem{explanation} The initial data are very small but
they do not vanish at the initial
point ($a=2.0001$ for electromagnetic waves  or $a=2.001$ for
gravitational waves) of the   null cone
 $\tilde \Gamma_a $ that is being defined in Secs IIIA and VA.
 Therefore the  integration along  $\tilde \Gamma_a$ gives a number that
is not
  strictly equal to $\delta E_a$  but still it constitutes
a valid approximation,  due to the smallness of initial data.
\end{references}
\end{document}